\documentclass[%
 reprint,
 amsmath,amssymb,
 aps,
]{revtex4-2}
 
\usepackage{graphicx}
\usepackage{dcolumn}
\usepackage{bm}
\usepackage{xcolor}
\usepackage{hyperref}
\usepackage{braket}

\begin{document}

\title{Scalable Fault-Tolerant Quantum Technologies with Silicon Colour Centres}

\author{Stephanie Simmons on behalf of Photonic, Inc.}

\date{\today}

\begin{abstract}

The scaling barriers currently faced by both quantum networking and quantum computing technologies ultimately amount to the same core challenge of distributing high-quality entanglement at scale. In this Perspective, a novel quantum information processing architecture based on optically active spins in silicon is proposed that offers a combined single technological platform for scalable fault-tolerant quantum computing and networking. The architecture is optimized for overall entanglement distribution and leverages colour centre spins in silicon (T centres) for their manufacturability, photonic interface, and high fidelity information processing properties. Silicon nanophotonic optical circuits allow for photonic links between T centres, which are networked via telecom-band optical photons in a highly-connected graph. This high connectivity unlocks the use of low-overhead quantum error correction codes, significantly accelerating the timeline for modular, scalable fault-tolerant quantum repeaters and quantum processors.

\end{abstract}

\maketitle


\section{Context}

Quantum information processing unlocks novel technological capabilities that cannot be achieved classically. To this end, there is a long and growing list of quantum algorithms~\cite{zoo}, to be executed on future quantum computers, which offer super-polynomial speedups over their known classical counterparts. Some of these algorithms are critical for useful tasks including quantum chemistry and the design of novel materials as they enable precise simulations of chemical processes in important technological applications ranging from catalysts~\cite{Troyer.2021} to batteries~\cite{Lee.2022} and pharmaceuticals~\cite{Rubin.2022}. Other quantum algorithms will impact our cybersecurity standards. For instance, once implemented on a large-scale fault-tolerant quantum computer, Shor's algorithm will allow efficient decryption of any data protected by RSA cryptography and its variants, which currently underpins $>$90\% of all financial and internet traffic~\cite{rsa90}. Interestingly, numerous so-called ``quantum-safe'' RSA-replacement encryption algorithms have already fallen to classical or quantum attacks, including Diffie-Hellman~\cite{Mariyappn.2020}, Soliloquy~\cite{Shepherd}, Elliptic Curve~\cite{Hensinger.2022,Sangouard.2023}, Rainbow~\cite{rainbow}, SIKE~\cite{sike}, and (through a side-channel) CRYSTALS-Kyber~\cite{crystals}. 
Some PQC algorithms under consideration have held up against attack so far, and have the potential for strong long-lasting security in practice, but similar to RSA, rest upon intrinsically unprovable computational security~\cite{PQC_review}.

In parallel to quantum computing, quantum networks provide greater connectivity and security for quantum devices, and enable applications beyond the power of stand-alone quantum devices. These include secure communications~\cite{BB84} that even future quantum computers cannot break, “blind” computing of both classical and quantum algorithms~\cite{Kashefi.2009}, and modular quantum computing~\cite{Cacciapuoti.2022}, timing~\cite{Lukin.2014}, and sensing~\cite{Zhuang.2021}. Analogously to today's high performance modular classical supercomputers, modular quantum supercomputers over quantum networks will enable astonishing computational capabilities by enabling true horizontal scaling of quantum resources, realizing practical quantum advantage for the aforementioned applications and giving rise to quantum applications yet to be imagined. Importantly, modular quantum processors connected by a quantum network directly offer all capabilities required of quantum repeaters, which are necessary to scale quantum networks' topologies and ultra-long distance quantum networking applications. 

However, today’s noisy intermediate-scale quantum (NISQ) devices cannot be trusted to execute practical-scale networking and computational instructions beyond small instances of a few specific problems \cite{peruzzo2014,farhi2014,farhi2000}.  
For each of today’s monolithic quantum architectures it has been a challenge to deliver high-fidelity operations at a scale of even a few hundred physical qubits, and presently it seems that more (and perhaps substantially more) high-fidelity qubits will be necessary for commercial advantage. 
Similarly, the scaling of quantum networks is currently blocked by the unavailability of reliable quantum repeaters. Therefore, it remains an open question if any commercial value can be realized with these NISQ devices~\cite{Kwek.2022}. 

Delivering the full potential of quantum information processing requires the construction of scalable fault-tolerant quantum (SFTQ) computing and networking technologies. Although the global race to deliver SFTQ technology is already considerable and accelerating, and the performance of some small-scale quantum systems are approaching the levels needed to operate individual logical qubits \cite{wang2023}, scaling such systems seems to be a considerable challenge, and hence the advent of large-scale SFTQ networking and computing technologies is believed by some to be a decade or more away~\cite{evolutionq} based upon the quantum architectures known today. 

\begin{figure*}[t]
\includegraphics[width=\textwidth]{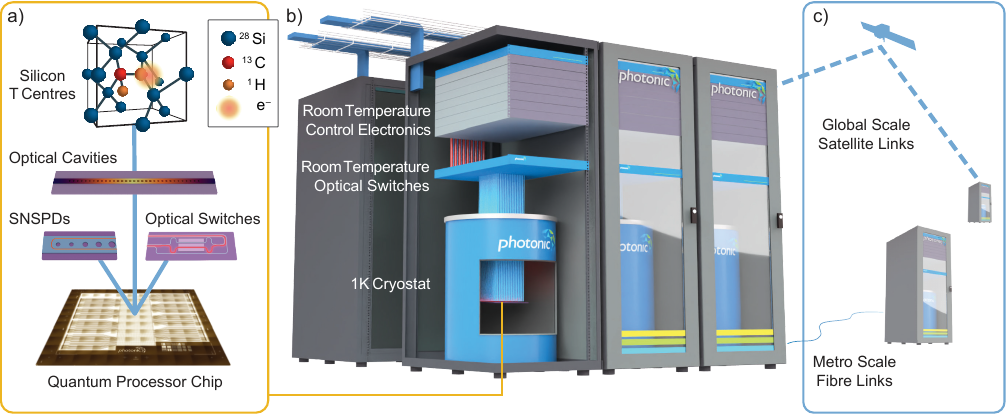}%
\caption{\label{fig:architecture} Photonic's scalable quantum technology architecture. A quantum chip is cooled in a 1~K cryostat. This chip hosts integrated silicon T centres within optical cavities, photonic switches, and single photon detectors. Optical input-output (IO) ports via telecom fibre connects to a room temperature photonic switch network and control electronics. This naturally allows a highly-connected architecture with non-local connectivity even as the system scales in size. Telecom fibre also enables horizontal system scaling by connecting multiple cryostats together through their optical IO. This enables both expansion of computing power and long-distance quantum networks.}
\end{figure*}

In this Perspective the argument is made that quantum networks and quantum information processors will both achieve ultimate scale by combining them into the same core entanglement distribution technology. The networking of quantum computers will allow for true horizontal scaling of quantum resources through modularity; and the introduction of quantum repeaters (small quantum computers) will scale telecom quantum networks' topologies and total distance. By way of illustration, a quantum architecture emphasizing entanglement distribution is proposed (Figure~\ref{fig:architecture}), which could be reasonably expected to deliver SFTQ technologies and the corresponding transformational real-world use cases far earlier than current predictions.

This technology is being built based upon networked spins in silicon, specifically the T centre spin-photon interface \cite{Simmons.2020}. The architecture itself is broadly applicable to all long-lived spin-photon interfaces such as solid state colour centres~\cite{B..2018} as well as some optically active atoms in vacuum ~\cite{Monroe.2004,Rempe.2007}. This architecture exploits the connectivity and modularity of quantum networks to scale the power of fault-tolerant quantum processors, and the error-corrected memory of distributed processors to scale the reach of quantum networks. This document is an overview of that vision.

\section{Connectivity}

The stringent thresholds for fault-tolerant quantum error correction imply that qubits must operate in strictly controlled environments, including (for example) low temperatures, ultra-high vacuum, electromagnetic shielding, with high-purity materials, and more. Each of these physical constraints result in a certain system size beyond which the marginal difficulty of adding the next qubit gets harder, not easier. Constraints such as these imply that, for each quantum computing platform, there exists a natural system size past which it would be far easier to link multiple computer modules instead of building ever-larger monolithic quantum supercomputers. This shift into modular quantum processing has multiple advantages: it can directly unlock quantum repeaters and scalable quantum networks if telecom photons are used to link the modules. For some quantum computing architectures the maximum number of future qubits in any one module could in principle be quite large, and the requirement to shift into horizontally-scalable modular quantum computing may be positioned in the distant future. However, for competitive (rapid) scaling of quantum resources, it is suggested in this Perspective that modular quantum processing will be an ultimate long-term goal for all quantum architectures on a practical basis.

If one assumes modularity will unleash limitless horizontal scaling of fault-tolerant quantum networks and quantum supercomputers, as it has for classical networking and supercomputers, the role of connectivity between modules deserves specific attention. For reference, modern classical high-performance computing hinges entirely upon parallelization across interconnected computing modules \cite{Gustafson1988,Asanovic2006}. In the quantum case even higher degrees of inter-module connectivity will be critical. Even with quantum algorithms that minimize the number of logical operations across modules \cite{Van.2016,Ferrari.2021,Cuomo.2023}, entanglement will need to be distributed efficiently. This is deeply connected to how quantum error correction is implemented in physical systems as is described next.

The scaling of connections between modules is a critical design parameter, as module size sets a lower bound on the number of entangled pairs that must be shared to apply an arbitrary operation on a joint two module system \cite{Stahlke.2011}. More importantly, connectivity remains significant in the specific case where each module encodes its own set of logical qubits. To implement logical two-qubit gates between separate modules, many physical qubits must be entangled between the individual modules. By way of example, for a CSS code, a transversal logical CNOT gate between logical qubits from two code blocks in separate modules requires transversal (pairwise) physical CNOT gates between every single one of their constituent physical qubits. Without a high degree of inter-module connectivity, the entanglement distribution between the two modules becomes a bottleneck in the performance of the distributed system~\cite{Nigmatullin2016, Simon2016}. As systems scale, this bottleneck can be avoided only if the connectivity of qubits between modules scales with the error-detecting capacity of the code. In codes where this error-detection capacity scales proportionally with module size, this poses a stringent mandate: an interconnect for each physical qubit in the ideal case~\footnote{As a prototypical example, consider the case where each module accommodates a codeblock of a large $n$-qubit error correction code with distance $d$. Here it is argued that a fault-tolerant entangling logical operation between these modules will always require $O(d)$ simultaneous physical entangling operations between the two modules. Performing entangling logical operations between two codeblocks necessarily involves at least $O(d)$ physical entangling operations, as the support of a logical Pauli in each module must grow by $O(d)$ in the other module, though these could be mostly within a codeblock in principle. If this operation is performed fault-tolerantly, any one single-qubit Pauli error cannot grow a support larger than $O(1)$ in either module. Since a logical Pauli operation in one module consists of the tensor product of $O(d)$ single qubit Pauli operations, $O(d)$ single qubit Pauli operations in one module must each gain distinct support from one another in the other module. This implies that there needs to be $O(d)$ entangling operations \emph{across} the two modules. Since we would like the total circuit depth for the logical operation to be as close to $O(1)$ as possible to maintain fault-tolerance, we should be capable of performing these $O(d)$ module-to-module entangling operations simultaneously. For example, this is how lattice surgery will be accomplished in planar QEC architectures. More importantly, if we consider an optimally scaling code where $d=O(n)$, this already implies we need to be able to apply $O(n)$ entangling operations in parallel between two $n$-qubit modules. A classic example of a logical operation with this requirement is the transversal CNOT operation in CSS codes.}. In large-scale quantum systems the interconnectivity of modules may in fact dominate the temporal and spatial quantum resource estimates of quantum networking and computing applications---and yet connectivity between and within modules is almost never accounted for, and typically not even mentioned, in current quantum resource estimation literature. Simply put, these one-to-one physical entangling operations between the modules must be parallelizable for good overall system performance. This observation implies certain physical qubit capabilities in the ideal case; it is argued below that maximal parallel entanglement distribution in a modular architecture preferably implies direct telecom optical access to each physical qubit. 

\begin{figure}[ht]
\includegraphics[width=\columnwidth]{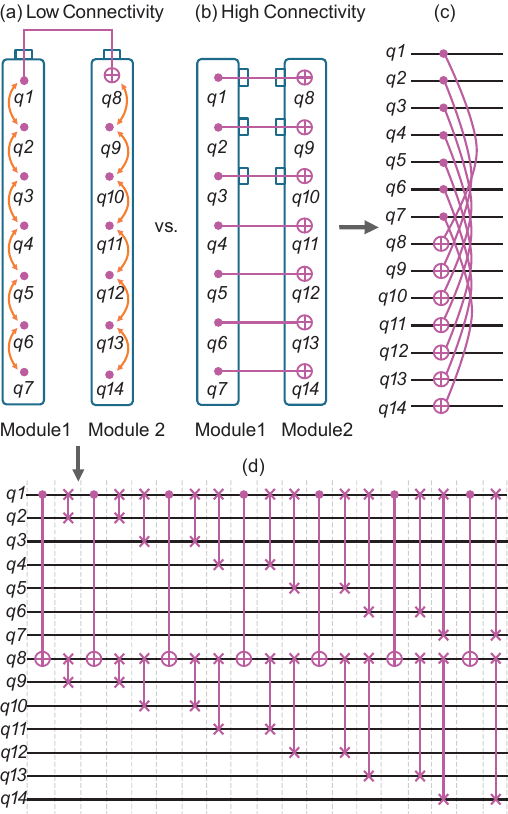}%
\caption{\label{fig:newfig} Temporal overheads associated with inter-module connectivity. An illustrative example of two 7-qubit CSS codeblocks with a) limited and b) ideal inter-module connectivity. c) Maximal inter-module connectivity where each physical qubit can be directly entangled with its partner qubit from the second module in parallel results in a single-timestep transversal CNOT gate, which implements the logical CNOT gate between these two modules. In d), the limited inter-module connectivity of this example results in a serial slowdown for the two-qubit logical CNOT operation with an illustrative, but sub-optimal, circuit implementation. Substantial additional temporal overheads are incurred (not shown here) if all-to-all connectivity within the module is not available as is assumed for this example. }
    \end{figure}

With the assumption of modular quantum technologies and high connectivity between modules for the purpose of high-performance parallel physical operations across modules, the natural question becomes: how should all of the qubits be connected to each other? The key to fast logical gates across modules is efficient nonlocal entanglement distribution. If indeed each qubit is designed to be easily entangled with other physical qubits from other modules, presumably that qubit supports a physical process which allows for nonlocal, as opposed to proximity-based, entanglement generation. Under such assumptions, connectivity both within and across modules are not necessarily constrained to proximity-based qubit topologies and can be connected in whatever way offers the best total system performance.

Most of today’s quantum architectures are built on the paradigm of planar proximity-based entanglement generation, and aim to leverage a variant of the well-studied low-connectivity quantum error correction (QEC) code known as the surface code~\cite{Kitaev.1998,Kitaev.2003} to achieve fault tolerance. Even these near-optimal codes for planar architectures require tremendous development and resources. For example, under optimistic settings for the factoring algorithm in Ref.~\cite{Ekera.2021}, the surface code would require upwards of 3000 physical qubits for every fault-tolerant logical qubit. Moreover, in many planar architectures, the vast majority of logical instruction cycles require swapping qubits in order to implement proximity-based multi-qubit operations~\cite{Gidney.2023}.

\begin{figure}[ht]
\includegraphics[width=0.8\columnwidth]{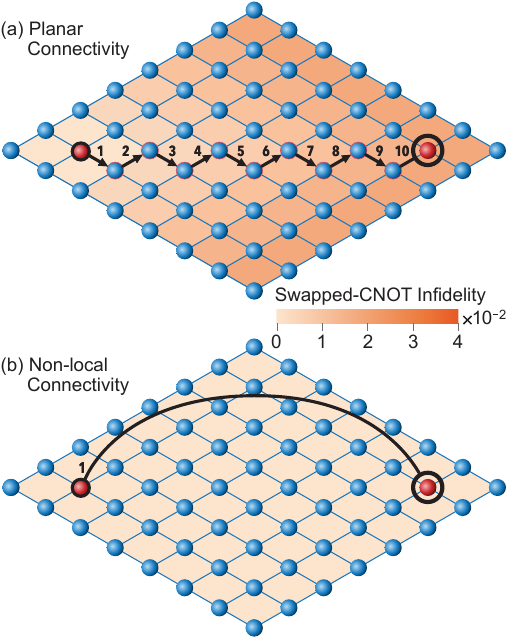}%
\caption{\label{fig:connectivity} Comparison between two-qubit operations on devices with (a) planar, nearest neighbour connectivity graphs and (b) non-local connectivity. A CNOT gate between two distant qubits on the planar graph requires sequential swap operations. Each consecutive operation accumulates errors and the overall fidelity drops precipitously, even over the small distances shown. With non-local connectivity, two qubit operations are equivalent across the graph and the device can scale with greater fidelity.}
    \end{figure}

Rejecting the assumption of planar connectivity (Figure~\ref{fig:connectivity}) for high-connectivity SFTQ architectures can give rise to a significant reduction of time and physical resource requirements~\cite{Vuillot.2017}. Avoiding these and similar overheads by allowing for higher connectivity can significantly reduce the number of operations required~\cite{Nickerson.2022,Sangouard.2023}. 
Even the distillation of magic states, which is largely the preferred method for providing the missing element of universality needed for surface code architectures, would be $2\times$ more resource-efficient using high-connectivity distillation algorithms~\cite{Haah.2012}, and near-term codes could have $10\times$ lower overhead~\cite{Brown.2022}. Unfortunately, neither a moderate amount of long-range connectivity~\cite{Beverland.2022}, nor small modules of qubits with substantial internal connectivity, offer a significant improvement over simple planar connectivity. The connectivity must keep pace with the system size to realize substantial improvements~\cite{Krishna.2022}.

Quantum Low Density Parity Check (QLDPC) codes, quantum analogues of the classical LDPC codes underpinning 5G telecommunications networks, are a class of codes characterized by low-weight stabilizer checks. When these codes are unshackled from the constraint of local connectivity, they can achieve truly astounding performance. A recent flurry of work~\cite{Tillich.2013,Hastings.2021,Breuckmann.2021,Panteleev.2021,Leverrier.2022,Panteleev.2022} culminated in the construction of QLDPC code families with asymptotically optimal code parameters, closing this long-open question after decades of relatively stagnant progress. In conjunction with previous work~\cite{Gottesman.2013}, these code families established that physical overheads could be driven significantly lower than those observed for planar code architectures. The inherent properties of QLDPC codes unlock additional performance perks including single-shot operation~\cite{Londe.2020,Quintavalle.2021,Higgott.2023,Gu.2023}, which removes the requirement to repeat QEC cycles to separate measurement errors from qubit errors \cite{Bombin.2015,Campbell.2019}. The work to implement logical operations has only just begun, but early signs point to methods on par to those of planar codes \cite{Quintavalle.2022,Cowtan.2023}. Surprisingly, no-go theorems implying the need for costly methods to achieve universality may not even apply to certain QLDPC codes~\cite{Bartlett.2022}.

Even more encouraging than these theoretical results on high-connectivity QLDPC codes has been the assessment of their practical performance, which appears to match or outperform the planar code in essentially every way. They offer similar error thresholds~\cite{roffe2020,Beverland.2022}, similar check weights~\cite{Brown.2022}, and proven efficient decoding algorithms for large finite codes~\cite{Kalachev.2021}. QLDPC codes that encode 1 logical qubit per every 3 physical qubits have been demonstrated to saturate the quantum hashing bound under depolarizing noise (one of the biggest points of pride of the planar surface code) using decoders already in practical use today \cite{yang2023spatiallycoupled}. Even with rather small system sizes (where the asymptotic behavior of QLDPC codes need not be representative of actual performance) and under nuanced noise models, QLDPC codes have been demonstrated to lower resource costs by over a factor of 10 in comparison to planar codes \cite{Bravyi.2023} even when extending to physical resource estimation \cite{Xu.2023}. Certain families of QLDPC codes offer constant overheads as the code scales in size. For instance, explicit codes of around 1,000 physical qubits support well over 100 logical qubits \cite{Xu.2023}. With constant overhead QLDPC codes, the encoding rate could remain the same as the system scales horizontally. Unfortunately for purely planar architectures, there is no substitute for high connectivity -- it is a fundamental requirement for any system hoping to capitalize on this tantalizing potential~\cite{baspin2022,delfosse2021}. While full spacetime logical circuit overhead estimates are currently scarce, existing methods~\cite{gottesman2013fault,Brown.2022} already upper bound this overhead at $O(n)$, and there is optimism that QLDPC logical circuit compiling can match or beat state-of-the-art methods for surface codes~\cite{beverland2022surface} in practice.

Even in advance of SFTQ, connectivity is also proving critical in NISQ applications. For example, to build a quantum repeater, connectivity allows for substantially more efficient entanglement distribution across many users~\cite{Zhang.2021,Englund.2022nr} and higher entanglement throughput via multi-path routing~\cite{Pirandola.2022}. 
Other examples where connectivity improves near-term applications include mapping circuits to hardware~\cite{Wille.2022}, and quantum image processing~\cite{Schladitz.2023}. 

Taken together, the rate and quality of nonlocal entanglement generation should be seen as the key characteristic dictating the ultimate performance and resource requirements of applications using large-scale modular fault-tolerant quantum systems. This is true for both large-scale networks leveraging quantum repeaters as well as modular and horizontally scalable quantum supercomputers. In this vision, quantum supercomputers and quantum networks are both ultimately large-scale entanglement distribution systems. Although many physical qubit types can in principle support nonlocal entanglement generation, and the architecture proposed here could be applied to each, practical details which impact the rate and quality of a qubit's nonlocal entanglement generation capabilities can have truly dramatic consequences on the final system performance, and any given physical qubit type should be selected for further development with this long-term requirement in mind.

\section{Building blocks: Silicon, Telecom, Memory, Entanglement} \label{sect:BB}

\subsection{Silicon}
Silicon is a pinnacle material for both quantum and classical applications. It is the industrial standard for high-performance integrated electronics, as well as for low-loss, high-density, photonic integrated circuits both active and passive. Spin qubits within silicon have also proven to be exceptional quantum memories---they have set performance records for fidelity~\cite{Morello.2015} and lifetimes~\cite{Thewalt.2013}. The industrial dominance and extensive development of silicon offers such incomparable competitive advantages that, historically, if a solution is found using silicon, the silicon solution usually wins.

\subsection{Telecom photons}
Telecommunications-band (telecom) photonic flying qubits will be the backbone of any highly connected global quantum network and the backbone of modular quantum computers. Telecom photons can be flexibly routed with arbitrary connectivity to connect matter qubits both locally and remotely, with low loss in cryogenic-compatible waveguides and at room temperature using modern telecommunications infrastructure. Although a number of research efforts are now advancing the transduction of other qubits into telecom photons~\cite{Painter.2020,Fink.2020,Lanyon.2019,Hanson.2022058, delaney2022}, the overhead of transduction processes can be avoided entirely by working with quantum systems that interact with telecom photons directly---such as the silicon T centre. In this Perspective, the claim is made that telecom photons are essential for high-connectivity quantum technologies at scale but are likely not practically sufficient on their own without a quantum memory.

\subsection{Quantum memory}
The central challenge with photonic qubits of all wavelengths, across quantum computing and quantum networking, is that they suffer from unavoidable loss. Each photonic process over optical links succeeds only a fraction of the time, and this is true even for telecom photons which have the lowest loss photonic components available by far (e.g. switches)~\cite{alibart2016}. In large-scale, high-connectivity settings, each switch layer introduces yet further loss. Quantum memories offer a straightforward path to protect against photon loss for both high-fidelity computing and networking applications. 

Quantum memories are necessary for the construction of quantum repeaters with the highest functionalities~\cite{azuma2022}. In repeaters, ultra-long quantum memory lifetimes are essential for high-fidelity qubit storage and the coordination of photon signals with long time-of-flight (long distance) connections, as well as in high-loss environments where many entanglement-generation attempts are necessary on average before photons successfully arrive. Solid-state colour centre qubits with access to a spin degree of freedom~\cite{B..2018,Hanson.2021s3}, as well as some optically active atoms in vacuum~\cite{Monroe.2004,Rempe.2007,Nadlinger2022}, offer all of the aforementioned ideal quantum functionality. Each of these physical qubit types offer a direct high quality photon interface into at least one long-lived~\cite{Thewalt.2013}, high-fidelity~\cite{Laucht.2022}, universally-controllable spin qubit~\cite{Wrachtrup.2014}. Every physical qubit in this category has a direct and dedicated photonic interface allowing for parallel entanglement generation. 

For high-connectivity quantum computing, a similar architectural advantage is proposed here where the photons distribute entanglement but do not process quantum information directly: the processing is done within the spin qubits. This design is inherently tolerant to photon loss as probabilistic entanglement-generation attempts can be repeated until success, without losing the spin qubit state~\footnote{Alternatively, the photons could themselves be considered as qubits to directly mediate nonlocal interactions through strong coupling schemes ~\cite{knaut2023}, which could be made loss-tolerant under a broker-client~\cite{benjamin2006} model of entanglement distribution.}. Essentially, the high connectivity between matter qubits -- which could be physically arranged on planar chips -- is provided by entanglement carried by telecom photons and the arbitrary connectivity that these photons provide. 

Maximally entangled Bell pairs (BPs - pairs of qubits in one of the four Bell states) of long-lived spins can be produced via a variety of photonic methods (see Figure~\ref{fig:gate} below for a specific implementation~\cite{Kok.2005}). Once entanglement in the form of BPs is delivered to the spins, this entanglement can be used as a resource to construct cluster states to be consumed for computation within the measurement-based computing paradigm~\cite{Briegel.2003,Jozsa.2005} or teleported gates in the traditional gate model of quantum computing~\cite{Chuang.1999,Eisert_2000}. Both of the cluster state, and teleported gate models of quantum computation allow for blind-computing applications over a network~\cite{Kashefi.2009,Fitzsimons.2015}. Below the focus is on teleported gates as an example implementation. 

For this kind of quantum technology, quantum networks’ repeaters and quantum computing modules will be almost identical in their core construction, which reflects the fact that the key to efficient, high-performance large-scale quantum networking and quantum computing is high-bandwidth high-connectivity entanglement distribution. In the long term, it is suggested that this unified technology will outperform architectures with disparate networking and computing cores that require extra layers of quantum interfaces which will inevitably compromise entanglement distribution rates and fidelities.

\subsection{Computing and networking with T centres} 

Even in a loss-tolerant design, achieving efficient distribution of entanglement at scale is a critical performance metric for high-connectivity SFTQ. This further motivates the adoption of solid-state spin-photon interfaces, ideally telecom colour centres, which can be directly placed into low-loss integrated photonic circuits in silicon. Direct integration not only maximizes photon collection efficiency from every single qubit, but also enhances the photon emission rate and quality using routine photonic engineering, while the spins, particularly long-lived nuclear spins, can perform at the high levels necessary for low-overhead QEC.

The above considerations foretell a future where scalable fault-tolerant quantum networks and scalable fault tolerant quantum computers will be the same core technology, with long-lived spins embedded into silicon integrated photonics, directly interfacing with telecom photons. In particular, silicon T centre~\cite{Mudryi.1981,Simmons.2020,Simmons.2022} is proposed as the exemplary foundational quantum unit: in addition to direct telecom access, it contains one unpaired electron spin and up to three spin-1/2 nuclear spins (one hydrogen and two carbon atoms, see Figure~\ref{fig:architecture}), each of which serves as a high fidelity and long-lived qubit with good performance comfortably above 1~K~\cite{Simmons.2020} where substantial cooling power is available for large-scale modular processors. 

Solid-state spin performance is largely determined by the host crystal environment. High fidelity performance and ultra-long coherence times~\cite{Tyryshkin.2012, Thewalt.2013} are common to many silicon spin centres including T centres~\cite{Simmons.2020}. In particular, the capability to isotopically purify silicon to the nuclear spin-free $^{28}$Si isotope has shown nuclear spin coherence times of 3 hours~\cite{Thewalt.2013}, with fidelities above 99.9\%~\cite{Morello.2015}. T centres in bulk $^{28}$Si samples show spin coherence times ($T_2$ Hahn echo) of 2.1~ms for the electron and 1.1 s for the hydrogen nuclear spin; both spins have $T_1$ lifetimes far longer than a second~\cite{Simmons.2020}. 

The T centre spins almost uniquely~\cite{xiong2023} possess a direct telecom photonic interface in silicon. It can efficiently interact with a pump laser pulse and emit a spin-entangled O-band photon. A photonic cavity can change the photonic environment around the T and enhance the emission of the photon into desired optical modes through the Purcell effect \cite{purcell1946}. These modes are coupled to optical waveguides with well-defined modes, and therefore spin-entangled photons can be accurately (and with low loss) routed to their destination either through integrated photonic waveguides or by coupling into optical fibre. Optical fibres can connect T centres across multiple chips, enabling a naturally modular and horizontally-scalable architecture. Modular scaling can relieve IO density challenges and environmental constraints that emerge as systems grow in size.

\section{Quantum Operations}
The silicon T centre, only recently rediscovered for quantum applications in 2020~\cite{Simmons.2020}, offers all the ideal characteristics listed above to enable high-performance SFTQ in the near and long term. It merges the advantages of silicon fabrication and scalability, telecom emission, and long-lived spin memories into a quantum system uniquely adapted to a modular, high-connectivity architecture. Below it is described how the qubits of the T centre perform high-fidelity operations according to this vision. 

\begin{figure}[b]
\includegraphics[width=\columnwidth]{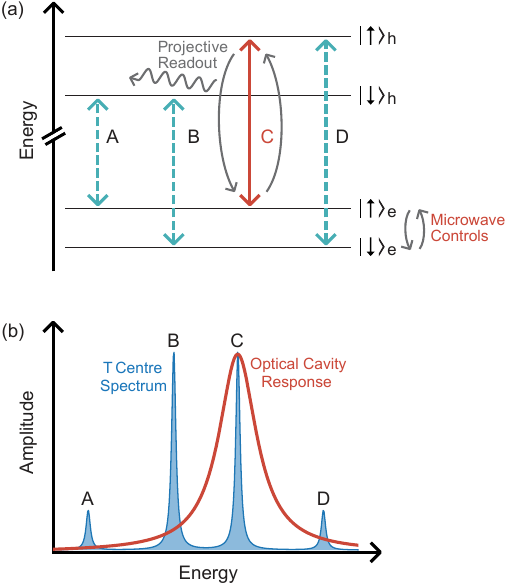}%
\caption{\label{fig:levels} (a) Energy levels of a T centre. Transition C is optically pumped to generate entanglement and for readout. h - excited state hole spin; e - ground state electron spin. (b) Optical spectrum of the T centre's electron spin, overlaid with optical cavity to enhance emission of the C transition.}
\end{figure}

\subsection{State preparation and measurement}
For readout and initialisation, the electron spin can be projectively measured and initialized to high fidelity using spin-dependent optical excitation (Figure~\ref{fig:levels}a)~\cite{Simmons.2020}. The initialization and readout of the full T centre spin register can be realized via successive rounds of high fidelity state swaps between the nuclear spins and the electron spin. Additionally, quantum non-demolition (QND) readout of the nuclear spins can be achieved by executing two-qubit CNOT gates with the electron spin as the target, optically reading out the electron spin, and then repeating as necessary~\cite{neumann2010}.

The four spin qubits in a T centre's local register admit independent and high-fidelity single-qubit and multi-qubit gates through standard electron and nuclear magnetic resonance techniques~\cite{Simmons.2020}. Additionally, the low spectral diffusion of the spin transitions in $^{28}$Si allows for full access to state-selective transitions, and high fidelity CNOT gates between the electron and its coupled nuclear spins are directly available~\cite{Simmons.2020}.

\subsection{Bell pair entanglement}

Two-qubit gates rely on proximity-based or nonlocal entangling operations between qubits. Remote T centres interact through photon-mediated entanglement. Each T centre can be optically triggered to emit single telecom O-band photons~\cite{Simmons.2022}, optionally entangled with the electron spin. To generate entanglement between spins, these photons must be indistinguishable. This means matching every degree of freedom between two photons: wavelength, linewidth, timing, polarization, and so on. Importantly, the wavelength of these photons can be tuned with either strain~\cite{safanov1995} or electric fields~\cite{Atature.2011,stark_forthcoming} to compensate for the variations in T centres’ local environments. Other degrees of freedom can be engineered to match using integrated photonics, and timing can be synchronized by calibrating control signals against optical path lengths. 

\begin{figure}[t]
\includegraphics[width=0.8\columnwidth]{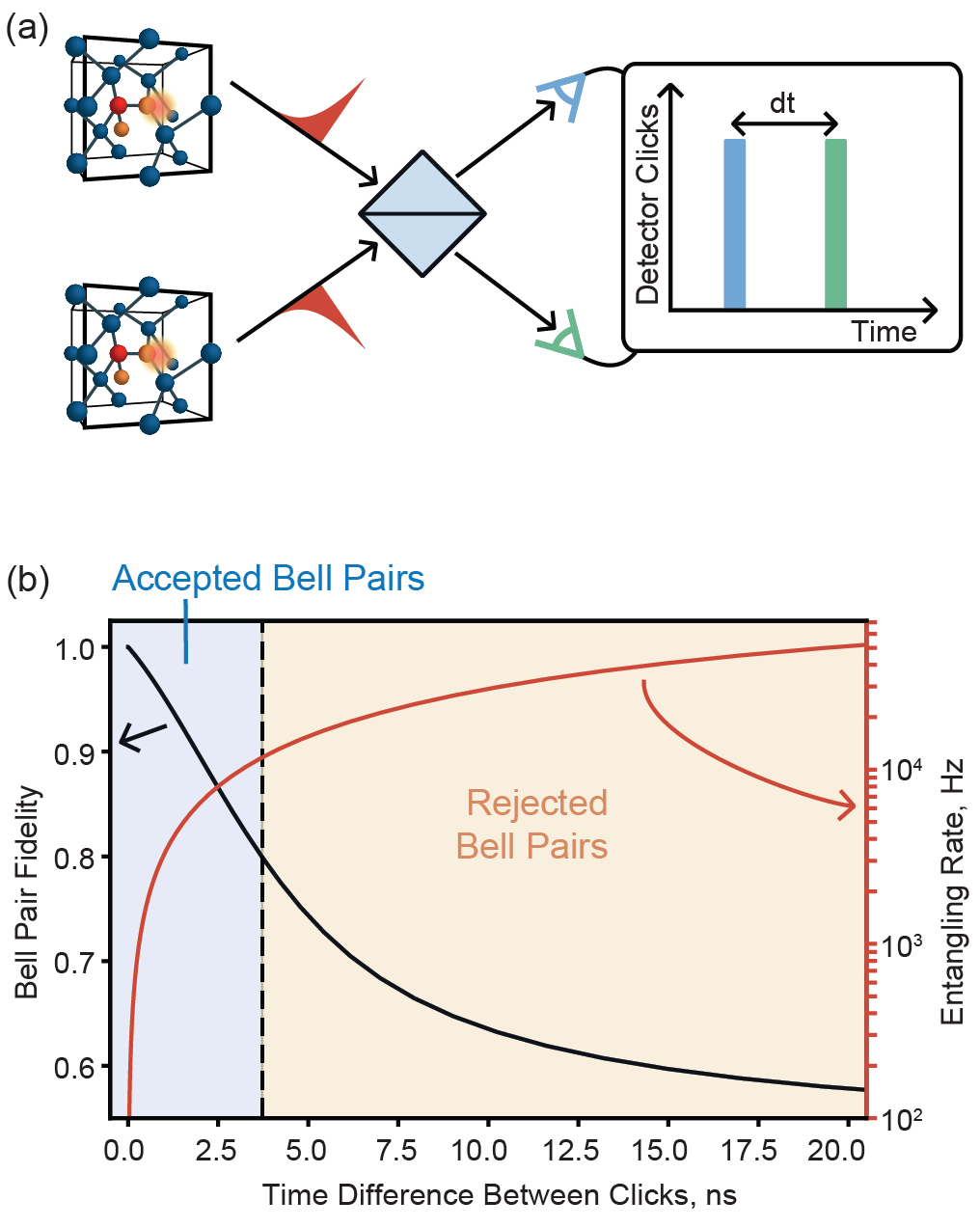}%
\caption{\label{fig:hom}(a) Two T centres emit synchronized photons. Detection times will vary due to the finite time width of the photon wavepackets. (b) The detection time difference $dt$ can be used to herald only high-fidelity Bell pairs. As the heralding threshold grows more stringent, the Bell pair generation rate decreases. The model curves above were calculated using reasonable projections of system parameters from current performance. }
\end{figure}

The indistinguishability of spin-entangled photons dictates the quality of the generated Bell pairs, and hence the maximum fidelity of the multi-qubit operations. Highly indistinguishable photons can be produced on demand when the emitter's optical transition linewidth approaches its lifetime-limited value, rather than being broadened by environmental noise. This can be dramatically assisted by photonic engineering, specifically by spin-selective Purcell enhancement of the optical transition when on resonance (Figure~\ref{fig:levels}b) with a high-Q photonic cavity~\cite{Awschalom.2020}. This offers faster emission rates, leading to larger lifetime contributions to the linewidth, and therefore more indistinguishable photons, as well as higher cyclicity (defined as the conservation of the spin state through an optical cycle).

Integration into photonic cavities has already led to demonstrations of $>20\times$ reductions in single T centre excited state lifetimes compared to the 940~ns unenhanced value~\cite{purcell_forthcoming}, enhancing indistinguishability and increasing maximum photon rates. Instantaneous linewidths for single T centres only $5\times$ larger than the lifetime-limited value have also been measured~\cite{purcell_forthcoming}. Additional improvements towards high indistinguishability include enhanced fabrication methods to reduce the optical spectral diffusion. 

The triggered emission of two indistinguishable high-purity spin-entangled photons can be used to project the participating electron spins into a maximally entangled Bell Pair spin state~\cite{moehring2007, Bernien_2013}. 
An illustrative example of the BP generation protocols that exist, the Barrett-Kok~\cite{Kok.2005} protocol family is notable in that it is which-way symmetric, does not require interferometric stability of the setup, and heralds success with minimal false positives.

\begin{figure}[t]
\includegraphics[width=\columnwidth]{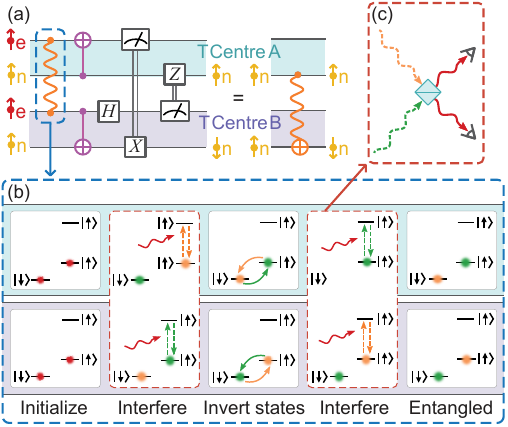}%
\caption{\label{fig:gate}Two-qubit gate mechanism. The electron spin is initialized in a superposition state and optically pumped to emit a spin-entangled photon. Repeating this process and heralding on photon detector clicks produces photon-mediated electron spin-spin entanglement, which is consumed in remote T centres to effect a CNOT gate between two nuclear spins.}
\end{figure}

The Barrett-Kok protocol proceeds as follows (Figure~\ref{fig:gate}): the electron spins of two separated T centres are prepared in a superposition of spin-up and spin-down, and triggered to emit a photon resonant with the spin-up transition. The photonic modes are interfered on a beamsplitter and detected. Then the spin states are inverted, and the emission and detection is repeated. Using exactly one photon detection in each of the first and second optical cycles as a herald signal, the spins are projected into a maximally-entangled BP. 

Although photon loss can cause an individual attempt to fail, success is heralded and the entire process can be repeated until success, with small degradation in the local nuclear spin coherence for each attempt (see below). Moreover, for this protocol, higher fidelity BPs can be obtained at the expense of lower entanglement rates by shifting the heralding threshold, the time difference between detector clicks, e.g. using time-bin filtering for time-frequency mismatch (Figure~\ref{fig:hom})~\cite{metz2008}.

\subsection{Teleported gates}

The teleported-gate model has a number of advantages at scale. Teleportation delivers two-qubit gates on demand, after coincident, identical photons are heralded. This inherently eliminates accidental multi-qubit operations and cross-talk that can occur when two-qubit interactions are ``always-on''.
Furthermore, qubits coupled to common local environments are susceptible to correlated errors that can dramatically degrade QEC performance~\cite{Wilen_2021, yoneda2022, rojasarias2023}. Since the teleported gate is a remote operation, each T centre can occupy a unique local environment, thus minimizing correlations in noise. 

Once the BP entanglement is heralded and delivered to the two T centres’ electron spins, that BP can be consumed in a teleported gate sequence to apply a non-local multi-qubit gate between the two T centres’ nuclear spins using only local operations and classical feed-forward operations (see Figure \ref{fig:gate})~\cite{Eisert_2000,Schoelkopf.2005,Neyenhuis.2021}. Alternatively, a series of successful BPs can be distilled through local measurements and gates onto a remote memory qubit BP ~\cite{Wootters.1996,Hanson.2017l5q}, with fidelity beyond the photon-indistinguishability, before this distilled pair is consumed to deliver higher fidelity teleported gates. A register of multiple spin memory qubits such as the T centre's internal nuclear spins allows a tiered distillation protocol \cite{Benjamin2012}. Distillation can prioritize the dominant error pathway of the teleported gate, yielding, for example, teleported CNOT process fidelity exceeding the heralded or distilled BP fidelity \cite{Bernien_2013}. 

The approach outlined here uses the electron as a spin-photon interface to generate and distribute entanglement, and nuclear spins for processing and memory. A configuration must be chosen that protects the memory qubits during entanglement distribution attempts. Repeated attempts can, in general, degrade the information stored in memory by perturbing the nuclear spin states \cite{Hanson.2021s3}. 

Protection techniques include working in a decoherence free subspace~\cite{Reiserer_2016}, aligning an external magnetic field along one of the principal axes of the hyperfine tensor~\cite{hyperfine_forthcoming}, and reducing the optical excited state lifetime~\cite{Jiang_2008, blok2015}. With sufficient protection, BP generation does not materially affect data stored in the nuclear spins and so the teleported gate operations are made photon-loss tolerant. Moreover, because of the long memory, this system works well for distillation and creating large entangled states, even after minimizing the required number of attempts by minimizing optical losses.

\section{Scalable quantum technology}

Scaling the number of interacting qubits in this architecture requires mediating entanglement across a growing network of qubits. Enabling multi-qubit operations between any two T centres amounts to routing their indistinguishable photons to the same beamsplitter and detector module. Here, the silicon-on-insulator (SOI) platform can be leveraged to drive on-chip device integration to a stage where photons never need to leave the photonic integrated circuit, maximizing photon collection efficiency and entanglement heralding rates. Furthermore, optical interconnects off-chip allow connections to outside networks and modular scaling of computing power.

A visual summary of this physical architecture is shown in Figure~\ref{fig:architecture}. T centres are placed into photonic cavities which are directly coupled into low-loss photonic waveguides. Emitted photons are routed on-chip through switches and, optionally, through high-efficiency optical IO ports into one of the many optical fibres. An optical switch fabric governed by run-time electronics guides T centre emission to single-photon detectors and Bell-state measurement modules within the cryostat.

The ability to then link distinct photonic chips through off-chip interconnects enables modularizing and distributing IO across multiple chips to horizontally scale the system size both within one and amongst many cryostats. Thanks to the low loss of telecom photons in optical fibre, the cryostats can be metres to kilometres apart.

\subsection{On-chip scalability}

T centres are atomic in scale and can be localized within small photonic components without direct optical crosstalk, making it possible to pattern millions of individually addressable qubits per chip~\cite{Simmons.2022}; however, the ultimate qubit count per chip will be governed by the larger footprint of active photonic components, control electronics, and signal IO. 

The low thermal conductivity of optical fibre enables incredibly high count optical IO into cryogenic environments (e.g. 37,000 optical fibres in Ref.~\cite{MacDonald_2015}).  High-density optical connections to integrated photonic devices have also been demonstrated in cryogenic environments~\cite{snspd8x8}. Furthermore, the T centre achieves operating performance (electron spin lifetime, optical linewidth) at only 1-2~K~\cite{Simmons.2020}. This relatively high operation temperature circumvents the need for dilution refrigeration technologies, and permits increased cooling power to counteract the heat load generated by the signal IO. As high density photonic IO continues to develop, it is expected that each chip could be able to support upwards of 4,000 fibres with 1,000+ qubits per system and beyond. 

 T centre qubits can be controlled using electrical signals segmented into global fields and local fields. Global fields include a static magnetic field $B_0$ to define the T centre spin Hamiltonian and a spin control field $B_1$ that can address qubits individually using global microwave control techniques~\cite{Yang.2021, Pla.2022, Kane.1998}. Local control fields include electric field $E$,  a trimming magnetic field $\delta B_0$, and optical switch and detector signals. Multiplexing strategies reduce the electrical line count for all signals, particularly the local control~\cite{Scappucci.2020} and detector~\cite{Siegel.2017} signals.

\subsection{Local quantum networks}

As mentioned in Section \ref{sect:BB}, quantum computing and quantum networking are essentially the same in the architecture presented here. The challenge that is overcome through this approach to SFTQ technologies -- high-bandwidth, high-connectivity entanglement distribution over lossy optical networks -- is the same challenge faced in scaling up quantum communication networks.

For this high-connectivity architecture, horizontal scaling of quantum resources is achieved by racking multiple units adjacent to each other into a modular local quantum network of qubits (Figure~\ref{fig:architecture}). Where additional network topology, inter-chip links, or connections between distant T centres are needed, optical IO can route photons to switch networks. Because optical photons can noiselessly traverse from cryogenics to room temperature without transduction, the switching network can be placed either within the cryostat or at room temperature. 

\begin{figure}[ht]
\includegraphics[width=\columnwidth]{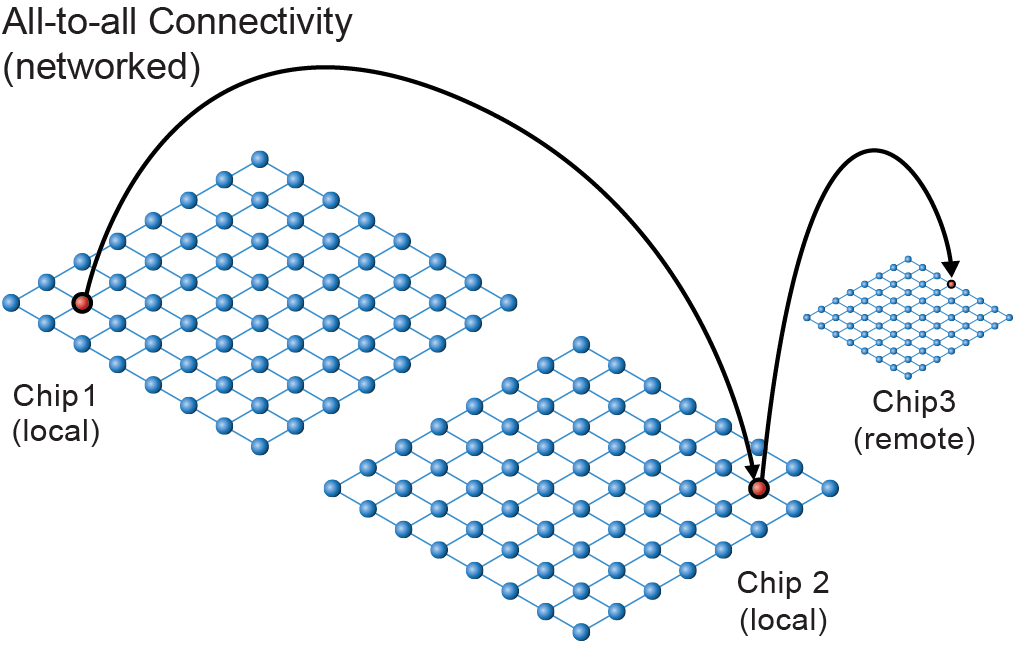}%
\caption{\label{fig:network connectivity} The non-local connectivity of the system works the same way when entangling qubits across the chips in the same cryostat, 10~m away, or 100~km away.}
\end{figure}

This design allows for arbitrary and flexible connectivity using switchable telecommunications hardware at room temperature with simple connections to cryogenics via optical fibre. Additionally, on-chip switching and multiplexing can then be used to scale the qubit count within each chip beyond limitations imposed by the IO fibre count. 

For the architecture described in this paper, T centres from distinct cryogenic units can be entangled in the same way as adjacent T centres (Figure~\ref{fig:network connectivity}). Put another way, the multi-qubit operations between different cryostats can be similar in performance to the multi-qubit operations within a chip. This design maximizes the performance of important distributed logical quantum algorithms~\cite{Troyer.20211clp}. The physical and logical qubits will have effectively all-to-all connectivity either by routing photons to common beamsplitters, by teleporting qubit states to new neighbourhoods of connectivity, or through entanglement swapping protocols~\cite{Cacciapuoti.2022}.

\subsection{Quantum repeaters}

Broad adoption of a quantum internet~\cite{kimble2008} is hampered by issues that arise when quantum communication channels are extended beyond local ($\approx 10~\mathrm{km}$) distances: the communication rate of point-to-point terrestrial fibre links drops exponentially with distance due to signal attenuation intrinsic in even the best available optical fibres. Similarly, beam divergence, atmospheric absorption, and turbulence impose fundamental losses and noise on free-space telescopic links.
And even though telecom networks have achieved point-to-point quantum communications over distances as large as $1000$~km \cite{liu2023}, optical quantum repeaters remain the key to unlocking global quantum networking applications~\cite{azuma2022}.

Fibre-coupled, T centre processor nodes contain telecom optical quantum memories that are capable of supporting long-distance communication over low-loss fibre links, and are able to generate, store and process  optically-entangled spin qubits.
Hence, they can be naturally operated as a telecom-wavelength quantum repeater to realize a long-distance, scalable multi-user quantum internet, with the potential to securely distribute entanglement or connect quantum processors between any two users on the global network (Figure~\ref{fig:architecture}).

As an initial implementation, the nested quantum repeater protocol~\cite{Briegel.1998} creates long distance entanglement by heralded entanglement generation over shorter links and entanglement swapping between adjacent nodes followed by entanglement distillation.
The steps are repeated between successive, now twice as long, entangled links until the end nodes are entangled. The individual entanglement generation steps taken during this procedure are exactly the same as local, distributed SFTQ computing using the proposed architecture, allowing implementation by T centre processors. This repeater protocol belongs to the so called first generation of quantum repeaters~\cite{muralidharan2016}. Here the loss errors are addressed by heralded entanglement generation and operational errors are addressed by entanglement distillation, which is a specific type of error correction.

Looking to the second generation of quantum repeaters, one can leverage the ability of T centre processors to perform multi-qubit encoding and logical operations between nodes. As before, loss errors on individual photons are addressed by heralded entanglement generation, but the operational errors are addressed by logical encoding at each node, namely with codes permitting transversal two-qubit operations~\cite{Lukin.2009}, i.e. the teleported CNOT implemented in Fig.~\ref{fig:gate}. Each node possesses one or more logical qubits, which become logically entangled to neighbouring logical qubits through entanglement distribution. Subsequent logical entanglement swapping allows the end-node logical qubits to be entangled, similar to the end-node physical qubits of the first generation of repeaters. This second generation uses the scalability and connectivity of T centres in individual nodes to distribute error-corrected logical qubits over long distance.

For these reasons, the proposed T centre based architecture is not only best suited to tackle the challenges of SFTQ computing, but also offers a very attractive platform for implementing large scale quantum networks and their various applications.

\section{Quantum networking applications}
A truly scalable network, extendable both in distance and number of users, unlocks powerful applications~\cite{Wehner.2018, Simon.2017}, such as quantum key distribution (QKD), distributed quantum computing, blind quantum computing~\cite{Kashefi.2009}, and enhanced sensing. Here, the focus is on two applications, blind quantum computing and quantum key distribution, and discuss how these could be naturally implemented in the proposed architecture.

\subsection{Memory-assisted, measurement-device independent quantum key distribution}
\begin{figure}[t]
\includegraphics[width=\columnwidth]{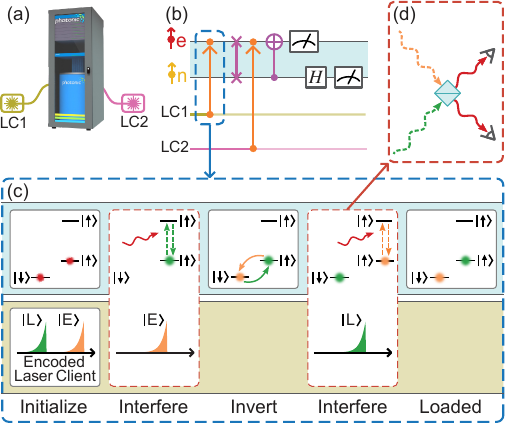}%
\caption{\label{fig:loading}(a) A protocol for loading of a time bin photonic qubit state into the spin memory of a T centre. (b) A quantum circuit that utilizes the protocol in (a) to generate a cryoptographic key for two users (laser clients, LCs) connected to the same hub (c). Loading photons from two users and then performing a Bell-state measurement on them computes the parity of their states, allowing secret key generation in MA-MDI QKD.}
\end{figure}

\begin{figure}[t]
\includegraphics[width=\columnwidth]{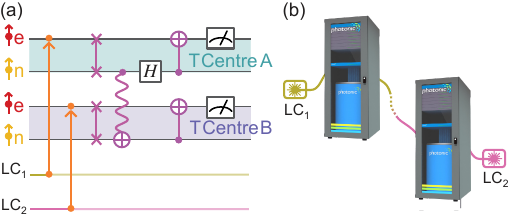}%
\caption{\label{fig:repeater} (a) Quantum circuit for loading qubit states from distant laser clients to two different quantum hubs and performing a nonlocal Bell state measurement between the two T centre spins. The teleported CNOT (middle) is mediated by entanglement distribution between the two hubs. (b) Because the CNOT between the two T centres is teleported, this circuit could be used to generate cryptographic keys between two users connected to different cryostats that are linked by telecom fibre. }
\end{figure}

Quantum key distribution is the quantum networking application nearest to widespread adoption. A first implementation of a QKD network using T centres consists of users possessing laser clients (LCs) that are optically connected to a quantum processor containing T centres (see Figure~\ref{fig:loading}), possibly in a hub-spoke configuration. Users can encode quantum information onto weak coherent pulses (for example, using time-bin encoding) and send this to the processor hub.
There, using the T centre's spin-photon entangled state and quantum teleportation, the encoded quantum state is loaded to a local spin state~\cite{Imamoglu.2013} for storage and further processing.
Multiple hubs could be interconnected to extend range and capacity (Figure~\ref{fig:repeater}).

The proposed networking configuration realizes memory-assisted measurement-device independent (MA-MDI) QKD~\cite{Lutkenhaus.2014}. Within this protocol a secret key between two users is established by first loading each user's states onto the hub's qubits followed by a Bell-state measurement between these two qubits. The result of that measurement is announced to the users, who use this information to extract a shared secret key.

The MA-MDI QKD protocol eliminates detector-side channel attacks~\cite{Qi.2012}, allowing the hub to operate as an untrusted node. Entanglement-based linking of nodes can also be untrusted, providing significantly more security than existing trusted-node networks~\cite{Zhang.2022}. At the same time, the memory-assistance provides improved keyrate scaling~\cite{Lutkenhaus.2014} due to the ability to store the clients' quantum information in the T centre spins. This means the client photons do not have to arrive simultaneously; their loading to the spins is heralded, and only once both are loaded is the Bell-state measurement performed between spins.

Within this architecture, each hub will support thousands of users in a hub-spoke model, well exceeding the usual point-to-point connections of currently commercially available QKD systems. Performing entanglement distribution between hubs as in Fig.~\ref{fig:repeater} allows implementing the same MA-MDI-QKD protocol in a distributed setting. The shared entanglement can be generated independently of the qubit loading events, and then consumed to implement the nonlocal Bell-state measurement. 

Whether connected to a single hub or a repeater chain, each user requires only a simple cost-effective source: a room-temperature laser attenuated below the single-photon level. It is equipped with a modulator able to produce time-bin encoded photonic qubits, which are spectrally matched to T centres in the hub. 
Such a device is very similar to a modern datacenter's telecom transceivers, and the photonic qubits produced by these sources can be similarly routed through existing datacentre or metropolitan telecom fibre to the hub.

\subsection{Blind quantum computing}

Although quantum repeaters allow for long-distance quantum computation and telecom-wavelength photons allow for modular quantum design, quantum computers may remain constrained by complex hardware required to protect and manipulate quantum states. This restriction suggests that quantum supercomputers are expected to start, and perhaps remain, as network-accessible large-scale devices. Cloud access over the classical internet may not be able to provide the necessary privacy for all potential end-user applications. Additionally, the operators of the quantum computers would have access to both the data and the algorithm run on it. Similar concerns for classical cloud computing galvanized research in the field of homomorphic encryption~\cite{acar2018}, where a computer performs logic directly on encrypted data without ever gaining the capability to decrypt and learn the underlying information.
Blind quantum computing~\cite{Kashefi.2009}, which may be viewed as a quantum analogue to homomorphic encryption, allows users to perform arbitrary computations using remote quantum processing resources, while restricting the computer from having meaningful access to both the data and the algorithm. In this highly-connected architecture, the same process of qubit loading employed for MA-MDI QKD can also be employed to realize remote and blind quantum computation, without a change of user hardware. This is due to the intrinsic spin-photon interface of the T centre, which allows loading arbitrary external user qubits into the computation in a heralded way. Loaded qubits are largely used to direct the computation remotely, while some are reserved as checks to confirm the operation of the computer~\cite{Ollivier.2022}.

\section{Conclusion}

The highest-value quantum applications known today require fault-tolerant capabilities \emph{at scale}. Modular quantum systems, potentially distributed over global distances, will be the ultimate version of horizontally-scaled quantum information processing and networking. Underpinning this view is the observation that combining quantum computing and quantum networking technology removes the fundamental obstacles to scale that each of these technologies are facing in isolation, ie truly scalable architectures are horizontally scalable. Namely, to unlock quantum networks at scale one needs to develop quantum repeaters which in the high-performance limit are essentially fault-tolerant quantum computing modules, and to unlock truly scalable quantum computing one needs to leverage the entanglement distribution capabilities of quantum networking to link quantum computing modules into quantum supercomputers. 

Given that high-bandwidth, high-quality entanglement distribution ultimately sets the performance of both scalable (modular) fault-tolerant quantum computing and networking, quantum systems should be engineered to optimize entanglement distribution. Under this model, quantum computing and quantum networking are (in the ideal case) ultimately the same fundamental technology.

This Perspective proposes a scalable quantum (networking \emph{and} computing) architecture with this end goal in mind. It suggests a specific implementation using telecom colour centres in silicon, namely the T centre, but this architectural model is broadly applicable to many qubit systems. Because of the high connectivity offered by the spin-photon interface, this architecture can take advantage of fixed- and low-overhead QLDPC codes to deliver fault tolerance. Thanks to the integrated silicon photonics platform, thousands of qubits can be fabricated and addressed on a single chip with optical and electronic control and routing, and modules can be connected together across existing global telecommunications infrastructure without any transduction losses. Using a T centre network to distribute verified quantum entanglement allows for device-independent networking protocols, providing the ultimate protection against eavesdropper attacks, as well as other high-value applications leveraging entanglement distribution such as blind computing. Taken together, the future for truly scalable quantum technology is bright.

\bibliography{references}

\end{document}